\documentclass[preprint,superscriptaddress,showkeys,showpacs,
tightenlines,fleqn,nofootinbib,prd,byrevtex]{revtex4} 
\usepackage{amsmath,amsfonts,amssymb,amscd,amsxtra,amsthm}
\usepackage{graphicx}
\usepackage{bm}
\begin{document}   
\preprint{INHA-NTG-09-04}
\preprint{CYCU-HEP-09-10}
\title{Photoproduction of the $\Theta^+$ and its vector and\\ axial-vector structure}      
%-------------------------------------------------
\author{Hyun-Chum Kim}
\email[E-mail ]{hchkim@inha.ac.kr}
\affiliation{Department of Physics, Inha University, Incheon 402-751, Republic
of Korea} 
%-------------------------------------------------
\author{Klaus Goeke}
\email[E-mail ]{klaus.goeke@tp2.ruhr-uni-bochum.de }
\affiliation{Institut f\"ur Theoretische Physik II, Ruhr-Universit\"
at Bochum, D44780 Bochum, Germany} 
%-------------------------------------------------
\author{Tim Ledwig}
\email[E-mail ]{ledwig@kph.uni-mainz.de}
\affiliation{Institut f\"ur Kernphysik, Universit\"at Mainz, D-55099 Mainz,
Germany} 
%-------------------------------------------------
\author{Seung-il Nam}
\email[E-mail ]{sinam@cycu.edu.tw}
\affiliation{Department of Physics, Chung-Yuan Christian University, Chung-Li 32023, Taiwan} 
%-------------------------------------------------

\date{\today}
\begin{abstract}  
We present recent investigations on the vector and axial-vector
transitions of the baryon antidecuplet within the framework of the
self-consistent SU(3) chiral quark-soliton model, taking into account
the $1/N_c$ rotational and linear $m_s$ corrections.  The main
contribution to the electric-like transition form factor comes from
the wave-function corrections. This is a consequence of the
generalized Ademollo-Gatto theorem. It is also found that in general
the leading-order contributions are almost canceled by the
rotational $1/N_c$ corrections. The results are summarized as follows: the vector and tensor $K^{*}N\Theta$ coupling constants,
$g_{K^{*}N\Theta}=0.74 - 0.87$ and $f_{K^{*}N\Theta}=0.53 - 1.16$, respectively, and $\Gamma_{\Theta \to KN}=0.71$ MeV, based on the result of the $KN\Theta$ coupling constant $g_{Kn\Theta}=0.83$.  We
also show the differential cross sections and beam asymmetries, based
on the present results. We also discuss the connection of present
results with the original work by Diakonov, Petrov, and Polyakov. 
\end{abstract}
\maketitle
%--------------------------------------------------
%--------------------------------------------------
\section{Motivation}
%--------------------------------------------------
We start with a brief summary of the present status about the
pentaquark baryon $\Theta^+$.  Since the LEPS collaboration announced
the evidence of the $\Theta^+$~\cite{Nakano:2003qx}, being motivated by
Diakonov et al.~\cite{Diakonov:1997mm} (DPP), there has been a great
deal of experimental and theoretical works on the $\Theta^+$  (see, for
example, reviews~\cite{Hicks:2005gp,Goeke:2004ht}).  However, the CLAS
collaboration reported null results of finding the
$\Theta^+$~\cite{CLAS1,CLAS2,CLAS3,CLAS4} in various 
reactions.  In this, we also want to mention the earlier work
\cite{Kwee:2005dz}. These null results from the CLAS experiment imply that
the total cross sections for photoproductions of the $\Theta^+$ should
be very small.  The KEK-PS-E522 collaboration~\cite{Miwa:2006if} found
a bump at around $1530$ MeV but with only $(2.5\sim2.7)\,\sigma$ 
statistical significance.  A later experiment at KEK (KEK-PS-E559),
however, has observed no clear peak structure for the $\Theta^+$ in
the $K^+p\to \pi^+ X$ reaction~\cite{Miwa:2007xk}. 

In the meanwhile, the DIANA collaboration has recently brought news on
a direct formation of a narrow $K^0p$ peak with mass of $(1537\pm  
2)$ MeV. The width of $\Gamma_{\Theta\to K^{0}p}=(0.36\pm0.11)$ MeV was also found~\cite{DIANA2}.  Compared to the former
measurement~\cite{Barmin:2003vv}, the decay width was more precisely
measured, the statistics being doubled.  The SVD experiment has also
announced a narrow peak with the mass,
$(1523\pm2_\mathrm{stat.}\pm3_\mathrm{syst.})$ MeV in the inclusive  
reaction $pA\to pK_s^0+X$~\cite{SVD,SVD2008}.  Furthermore, the LEPS
collaboration has reported again the evidence of the 
$\Theta^+$~\cite{Nakano:2008ee}: The mass of the $\Theta^+$ is found
at $(1525\pm 2+3)$ MeV and the statistical significance of the peak
turns out to be $5.1\,\sigma$.  The differential cross section was
estimated to be $(12\pm2)\, \mathrm{nb/sr}$ in the photon energy
ranging from 2.0 GeV to 2.4 GeV in the LEPS angular range.  In connection to the new LEPS results, Diakonov and Petrov recently 
discussed further thoretical aspects of the
$\Theta^+$~\cite{Diakonov:2008hc}.

Based on these experimental results, regardless of the existence of 
the $\Theta^+$ or not, one can come to the following three main
conclusions:  
\begin{enumerate}
\item The decay width of the $\Theta^+$ is very small ($\Gamma_{\Theta\to KN}<1$ MeV), which indicates that the $KN\Theta$ coupling should be
  tiny. 
\item The total cross section of the $\Theta^+$ photoproduction is
  small.  It implies that the $K^*N\Theta$ coupling constant should be
  also very small.  
\item Finding the $\Theta^+$ may be reaction-dependent. 
\end{enumerate}
We need to understand this smallness of the $KN\Theta$ and $K^*N\Theta$
coupling constants theoretically. In this, we will present results of recent investigations on the two coupling constants from the chiral quark-soliton model ($\chi$QSM)~\cite{Ledwig:1900ri,Ledwig:2008rw} and of the application of these results to the photoproduction of the $\Theta^{+}$~\cite{Nam:2009ds}. 

The present talk is schetched as follows: In Section 2, we review
briefly the general formalism of the $\chi$QSM to show how to
calculate the vector and axial-vector form factors. In Section 3, we
present the results and discuss them. We also predict the decay width
of the $\Theta^+$.  In Section 4, we discuss the connection of the
present reported results to the original work by DPP. In Section 5, we
describe the photoproduction of the $\Theta^{+}$ using the results from the $\chi$QSM. The last section is devoted to summary and conclusions
of the present reported work. 
%--------------------------------------------------
\section{Vector and axial-vector transition form factors}
%--------------------------------------------------
We start with the $\Theta^{+}$-to-neutron 
transition matrix elements of the vector and axial-vector currents
defined as: 
\begin{eqnarray}
&&\langle\Theta(p^{\prime})|\bar{s}\gamma^{\mu}u|n(p)\rangle = 
\bar{u}_{\Theta}(\bm{p}^{\prime})\left[F_{1}^{n\Theta}(Q^{2}) 
  \gamma^{\mu} + \frac{F_{2}^{n\Theta}(Q^{2})
    i\sigma^{\mu\nu}q_{\nu}}{M_{\Theta}+M_{n}} +
  \frac{F_{3}^{n\Theta}(Q^{2}) q^{\mu}}{M_{\Theta} +
    M_{n}}\right]u_{n}(\bm{p})\,\,\,,\label{eq:Vector-current}\\  
&&\langle\Theta(p^{\prime})|\bar{s}\gamma^{\mu}\gamma_5 u |n(p)\rangle 
=\bar{u}_{\Theta}(\bm{p}^{\prime}) \left[G_{1}^{n\Theta}(Q^{2})
  \gamma^{\mu} + G_{2}^{n\Theta}(Q^{2}) q^{\mu} +
  G_{3}^{n\Theta}(Q^{2}) P^{\mu}\right]\gamma^{5}\,
u_{n}(\bm{p})\,\,\,,\label{eq:Axial-Vector-current} 
\end{eqnarray}
where the $u_{\Theta(n)}$ denotes the spinor of
the $\Theta^{+}$ (neutron) with the corresponding mass
$M_{\Theta(n)}$.  The $Q^2$ stands for the momentum transfer
$Q^2=-q^2=-(p'-p)^2$ and $P$ represents the total momentum $P=p'+p$.
$F_i^{n\Theta}$ and $G_i^{n\Theta}$ stand for real transition form
factors that will be related to the strong coupling constants for the
$K^{*}N\Theta$ and $KN\Theta$ vertices with the help of the vector-meson dominance (VMD)~\cite{Sakurai:1960ju,Feynman:1973xc} and Goldberger-Treiman relation.  

In the VMD, the vector-transition current can be expressed as the
$K^*$ current by the current field identity (CFI):  
\begin{equation}
V^{\mu}(x) \; = \; \bar{s}(x) \gamma^{\mu}u(x) = 
\frac{m_{K^{*}}^{2}}{f_{K^{*}}}K^{*\mu}(x)\,,\label{eq:CFI}
\end{equation}
where $m_{K^*}$ and $f_{K^*}$ denote, respectively, the mass of the
$K^*$ meson, $m_{K^{*}}=892$ MeV, and decay constant defined as 
\begin{equation}
f_{K^{*}}^{2}=\frac{m_{K^{*}}^{2}}{m_{\rho}^{2}}f_{\rho}^{2}.
\end{equation}
The decay constant $f_\rho$ for the rho meson can be determined as 
\begin{equation}
f_{\rho}^{2}=\frac{4\pi\alpha^{2}m_{\rho}}{3\,\Gamma_{\rho^0\to e^+e^-}},
\end{equation}
where $\alpha$ denotes the electromagnetic fine-structure
constant. The $f_{K^{*}}$ is determined by the experimental data for
$\rho$-meson, $m_{\rho}=770$ MeV and $\Gamma_{\rho^0\to
e^+e^-}=(7.02\pm0.11)$ keV~\cite{Amsler:2008zzb}, for which we obtain the
values $f_{\rho}\approx4.96$ and $f_{K^{*}}\approx5.71$.

Using the CFI, we can express the $K^*N\Theta$ vertex in terms
of the transition form factors in Eqs.~(\ref{eq:Vector-current}) and
(\ref{eq:Axial-Vector-current}):  
\begin{eqnarray}
\langle\Theta(p^{\prime})|\bar{s}\gamma^{\mu}u|n(p)\rangle & = &
\frac{m_{K^{*}}^{2}}{f_{K^{*}}}\,\frac{1}{m_{K^{*}}^{2}-q^{2}}\,
\langle \Theta(p^{\prime})|K^{*\mu}|n(p)\rangle,\\ 
\langle\Theta(p^{\prime})|K^{*\mu}|n(p)\rangle 
& = &
\bar{u}_{\Theta}(\bm{p}^{\prime})\left[g_{K^{*}n\Theta}
  \gamma^{\mu} + f_{K^{*}n\Theta}
  \frac{i\sigma^{\mu\nu}q_{\nu}}{M_{\Theta}+M_{n}} +
  \frac{s_{K^{*}n\Theta} q^{\mu}}{M_{\Theta} + M_{n}} \right]
u_{n}(\bm{p}),
\label{eq:VDM2}  
\end{eqnarray}
where the $g_{K^{*}n\Theta}$ and $f_{K^{*}n\Theta}$ denote the vector 
and tensor coupling constants for the $K^{*}N\Theta$ vertex,
respectively.  These relations yield immediately the strong 
coupling constants as 
\begin{equation} 
g_{K^{*}n\Theta} \;=\;  f_{K^{*}} F_{1}^{\Theta n}(0),\;\;
f_{K^{*}n\Theta} \;= \; f_{K^{*}}\, F_{2}^{\Theta n}(0)\,.
\label{eq:gf}
\end{equation} 
Using the generalized Goldberger-Treiman relation, we can get the
strong coupling constant $g_{Kn\Theta}$ for the $KN\Theta$ vertex as
follows: 
\begin{equation}
g_{Kn\Theta}=\frac{G_{1}^{\Theta n}(0)\,(M_{\Theta}+M_{n})}{2f_{K}},
\label{GT_generalized} 
\end{equation} 
where $f_{K}\approx1.2f_{\pi}$ stands for the kaon decay constant.

In the rest frame of the $\Theta^+$, the form factors
$F_{1}^{n\Theta}(Q^{2})$, $F_{2}^{n\Theta}(Q^{2})$ and $G_{A}^{\Theta
  n}(Q^{2})$ of 
Eqs.~(\ref{eq:Vector-current},\ref{eq:Axial-Vector-current})  
can be expressed in terms of the matrix elements of the vector and
axial-vector currents with their time and space components decomposed 
in the $\Theta^+$ rest frame, respectively
\begin{eqnarray}
G_{E}^{n\Theta}(Q^{2}) & = & \int\frac{d\Omega_{q}}{4\pi} \langle
\Theta(p^{\prime})| \bar{s}\gamma^{0}u|n(p)\rangle,\cr
G_{M}^{n\Theta}(Q^{2}) & = & 3M_{n}\int \frac{d\Omega_{q}}{4\pi}
\frac{q^{i} \epsilon^{ik3}}{i\bm{q}^{2}} \langle
\Theta(p^{\prime})| \bar{s}\gamma^{k}u|n(p)\rangle\,\,\,,\label{eq:GM}\\
G_{1}^{n\Theta}(Q^{2}) & = & -\frac{3}{2{\bm{q}}^{2}}
\sqrt{\frac{2M_{\Theta}}{E_{\Theta}+M_{\Theta}}}
\int\frac{d\Omega_{q}}{4\pi} \Big[{\bm{q}} \times \Big({\bm{q}} \times
\langle \Theta(p^{\prime})| \bar{s}\bm\gamma\gamma_5 u
|n(p)\rangle\Big)\Big]_{z}\,, 
\label{eq:GA}
\end{eqnarray} 
where the electromagnetic-like Sachs form factors $G_{E}^{n\Theta}$
and $G_{M}^{n\Theta}$ are written as 
\begin{eqnarray}
G_{E}^{n\Theta}(Q^{2}) & = & F_{1}^{n\Theta}(Q^{2}) 
\label{eq:GE in Fi}\\
G_{M}^{n\Theta}(Q^{2}) & = & F_{1}^{n\Theta}(Q^{2}) + F_{2}^{n\Theta}(Q^{2})\,
\label{eq:GM in Fi}
\end{eqnarray}
The vector and tensor coupling constants are therefor obtained from
Eq. (\ref{eq:gf}) as:  
\begin{equation}
g_{K^{*}n\Theta} \;=\;  f_{K^{*}} G_{E}^{\Theta n}(0),\;\;
f_{K^{*}n\Theta} \;= \; f_{K^{*}}\, (G_{E}^{\Theta
  n}(0)\,-\,G_{M}^{\Theta n}(0)). 
\label{eq:gf2}
\end{equation}

We are now in a position to evaluate the form factors within the 
self-consistent $\chi$QSM. This model has the following virtues.
There are only three free parameters among which two are fixed in the 
mesonic sector and just one remains for the whole baryon sector. 
This allows to calculate the $\Theta^+$ transition form factors in the
same frame as was used for the proton electromagnetic form
factors.   
 
The model is featured by the following
effective low-energy partition function with quark fields $\psi$ with
the number of colors $N_{c}$ and the pseudo-Goldstone boson field
$U(x)$ in Euclidean space:  
%EQUATION>>>
\begin{eqnarray}
\mathcal{Z}_{\mathrm{\chi QSM}} &=& \int\mathcal{D} \psi\mathcal{D}
\psi^{\dagger}\mathcal{D}U\exp \left[-\int d^{4}x \psi^{\dagger}iD(U)
  \psi\right]= \int\mathcal{D}U\exp( -{\bf \mathcal{S}}_{\mathrm{eff}}[U]),
\label{eq:part}
\end{eqnarray}
\begin{equation}
\label{eq:echl}
{\bf \mathcal{S}}_{\mathrm{eff}}(U) =-N_{c}\mathrm{Tr}\ln iD(U),
\end{equation}
%EQUATION<<<
where
\begin{eqnarray}
D(U) & = & \gamma^{4}(i\rlap{/}{\partial} - \hat{m}-MU^{\gamma_{5}})
= -i\partial_{4}+h(U)-\delta m,
\label{eq:Dirac}\\
\delta m & = &
\frac{m_{s}-\bar{m}}{3}\gamma^{4}\bm{1}_{3\times3}+
\frac{\bar{m}-m_{s}}{\sqrt{3}} \gamma^{4} \lambda^{8}
= M_{1} \gamma^{4} \bm{1}_{3\times3}+M_{8}\gamma^{4}\lambda^{8}.
\label{eq:deltam}
\end{eqnarray}
The current-quark mass matrix is defined as
$\hat{m}=\mathrm{diag}(\bar{m},\,\bar{m},\, 
m_{\mathrm{s}})=\bar{m}+\delta m$.  The $\bar{m}$ stands for the
average of the up and down current-quark masses with isospin symmetry
assumed.  The $M$ denotes the constituent-quark mass of which the best
value for the numerical results is $M=420$ MeV.  The pseudo-Goldstone
boson field $U^{\gamma_5}$ is defined as 
\begin{equation}
U^{\gamma_5} \;=\; \exp(i\gamma_5 \lambda^a \pi^a) \;=\;
\frac{1+\gamma_5}{2} U + \frac{1-\gamma_5}{2} U^\dagger 
\end{equation}
with $U=\exp(i \lambda^a \pi^a)$.  For the quantization, we consider
here Witten's embedding of SU(2) soliton into SU(3):
\begin{equation}
U_{\mathrm{SU(3)}} \;=\; \left(
  \begin{array}{cc}
U_{\mathrm{SU(2)}} & 0 \\ 0 & 1     
  \end{array} \right)
\end{equation}
with the SU(2) hedgehog chiral field
\begin{equation}
U_{\mathrm{SU(2)}} \;=\; \exp[i\gamma_5 \hat{\bm n}\cdot \bm\tau
P(r)],   
\label{eq:hedgehog} 
\end{equation}
Here, the $P(r)$ denotes the profile function of the chiral soliton 
$U_{\mathrm{SU(2)}}$.  We refer to
Refs.~\cite{Christov:1995vm,Kim:1995mr} as to how one can 
compute the form factors within the $\chi$QSM.

%-------------------------------------------------
\section{$KN\Theta$ and $K^*N\Theta$ coupling constants}
%-------------------------------------------------
The results for the $K^*N\Theta$ and $KN\Theta$ coupling constant are
listed in Table~\ref{tab:coup}~\cite{Ledwig:1900ri,Ledwig:2008rw}.
\begin{table}[h]
\begin{tabular}{ccc|ccc}
\multicolumn{3}{c|}{$m_{\mathrm{s}}=0$}&
\multicolumn{3}{c}{$m_{\mathrm{s}}=180$ MeV}\\
\hline
$g_{K^*N\Theta}$ & $f_{K^{*}N\Theta}$ & $g_{KN\Theta}$&
$g_{K^*N\Theta}$ & $f_{K^{*}N\Theta}$ & $g_{KN\Theta}$ \\
\hline
$0$ & $2.91$ & $1.41$&
$0.81$ & $0.84$ & $0.83$ \\
\end{tabular}
\caption{The results for the $K^*N\Theta$ and $KN
\Theta^+$ coupling constants at $Q^2=0$ with and without
$m_{\mathrm{s}}$ corrections. The constituent quark mass $M$ is taken
to be $M=420$ MeV.} 
\label{tab:coup}
\end{table}

The stong influence of the $m_{\mathrm{s}}$ corrections on these
observables lies in a cancelation effect between the leading order and
$1/N_c$ corrections. This effect was already observed in the work of
DPP and led to the original prediction of the small $\Theta^+$ decay
width. 

Note that the vector coupling constant $g_{K^* n\Theta}$ vanishes in
exact SU(3) symmetry due to the generalized Ademollo-Gatto
theorem\footnote{We want to mention that the generalized
Ademollo-Gatto therem was first done by M.V. Polyakov.}, which will be
explained below.  The value of $g_{K^* n\Theta}$ with SU(3)
symmetry breaking comes solely from the wavefunction corrections, so
that the Dirac transition form factor turns out to be 
\begin{equation}
F_1^{n\Theta} (0) \;=\; \sqrt{3} c_{\overline{10}}^n
(1+\mathcal{O}(m_{\mathrm{s}})),  
\label{eq:ag}
\end{equation}
where $c_{\overline{10}}^n$ denotes the mixing parameter defined as 
\begin{equation}
c_{\overline{10}}^n \;=\; \frac{\langle n_{\overline{10}}|
  H_{\mathrm{sb}}|n\rangle}{M_n - M_{n_{\overline{10}}}}  
\end{equation}
with a symmetry-breaking part of the Hamiltonian $H_{\mathrm{sb}}$.
$M_{n_{\overline{10}}}$ denotes the mass of the antidecuplet neutron.
We call Eq.~(\ref{eq:ag}) as the generalized Ademollo-Gatto 
theorem. The results listed in Table~\ref{tab:coup} were obtained by using the
value of the constituent-quark mass $M=420$ MeV.  However, if we
calculate the coupling constants with $M$ varied from 400 MeV to 450
MeV, we get the vector and tensor $K^*N\Theta$ coupling constants as $g_{K^{*}N\Theta}=0.74 - 0.87$ and
$f_{K^{*}N\Theta}=0.53 - 1.16$, respectively. The smallness of these coupling
constants can be understood by comparing them with the $K^*p\Lambda$
coupling constants
\begin{equation}
|g_{K^*p\Lambda }| = 6.97,\;\;\;   |f_{K^*p\Lambda}| = 10.15,
\end{equation}
which were derived within the same framework.  Thus, the $K^*n\Theta$
coupling constants are indeed very tiny, which is in agreement with
the conclusion of recent experimental
data~\cite{Miwa:2006if,Miwa:2007xk,Nakano:2008ee}. Using the value of $g_{Kn\Theta}=0.83$. we immediately obtain the decay width of the $\Theta^+$ as $\Gamma_{\Theta \to KN}=0.71$ MeV which is in qualitative agreement with the data of the DIANA collaboration~\cite{DIANA2}.

The coupling constants for the proton can be obtained easily by
considering isospin factors.  Note that there is a sign difference in  
the coupling constants for the neutron and proton:
$g_{K^*n\Theta}=-g_{K^*p\Theta}$ and the same for the
$f_{K^*N\Theta}$~\cite{Ledwig:2008rw}.

%-------------------------------------------------
\section{Discussion of the value of the $\Theta^+$ decay width by
  Diakonov et al.}
%-------------------------------------------------
Diakonov et al.~\cite{Diakonov:1997mm} estimated the decay width of
the $\Theta^+$: $\Gamma_{\Theta\to KN}\approx 15$ MeV, based on the
experimental data for the $\pi NN$ coupling constant and 
\begin{equation} 
g_{\pi NN} \approx \frac{7}{10}\left(G_{0} + \frac{1}{2}G_{1}\right)\approx13.3,
\label{eq:gpNN}
\end{equation} 
where terms proportional to $G_2$ and $c_{\overline{10}}$ were
neglected.  However, the coupling constant $g_{Kn\Theta}$ is
proportional to $G_0-G_1$, so that it is not possible to determine it
by the $g_{\pi NN}$ only.  Thus, DPP~\cite{Diakonov:1997mm} have 
taken the results from the $\chi$QSM
calculations~\cite{Christov:1993ny,Blotz:1994wi}. A recent
work~\cite{Ledwig:2008rw} uses the same formalism as
Refs.~\cite{Christov:1993ny,Blotz:1994wi} but several parts
have been further elaborated. The symmetry-conserving quantization
\cite{Praszalowicz:1998jm} was established after the publication of
Ref.~\cite{Diakonov:1997mm}.  Since then, many observables of the 
baryon octet have been recalculated.  The quark densities of the
axial-vector current were also calculated~\cite{Silva:2005fa} and the
ratio $G_1/G_0$ can be found to be $0.68$.
    
Had DPP~\cite{Diakonov:1997mm} used the ratio 
$G_{1}/G_{0}=0.68$ instead of $G_{1}/G_{0}=0.4$, the decay width would
have turned out to be $\Gamma_{\Theta\to KN}=3.4$ MeV, which is much
smaller than the value $\Gamma_{\Theta\to KN}<15$ MeV published in 
Ref.~\cite{Diakonov:1997mm}, whereas $\Gamma_{\Delta\to\pi N}$ would remain
unchanged.  Even though we consider the criticism of
Ref.~\cite{Jaffe:2004qj} with $G_{1}/G_{0}=0.68$, one will get 
the decay width of the $\Theta^{+}$ which is also smaller than
predicted in Ref.~\cite{Diakonov:1997mm}.  Thus, we want to point out
that the predicted physics in Ref.~\cite{Diakonov:1997mm} by using the
$\chi$QSM is therefore unchanged.  
%-------------------------------------------------
\section{Photoproduction of the $\Theta^+$}
%-------------------------------------------------
The results of Ledwig et al.~\cite{Ledwig:1900ri,Ledwig:2008rw} enable
us to proceed to investigate the photoproduction of the $\Theta^+$
unambiguosly.  In Ref.~\cite{Nam:2009ds}, the $\gamma N\to K\Theta^+$
reaction has been revisited.  We will briefly review several results
of Ref.~\cite{Nam:2009ds} in this Section. 
\begin{figure}[t]
\includegraphics[width=7.5cm]{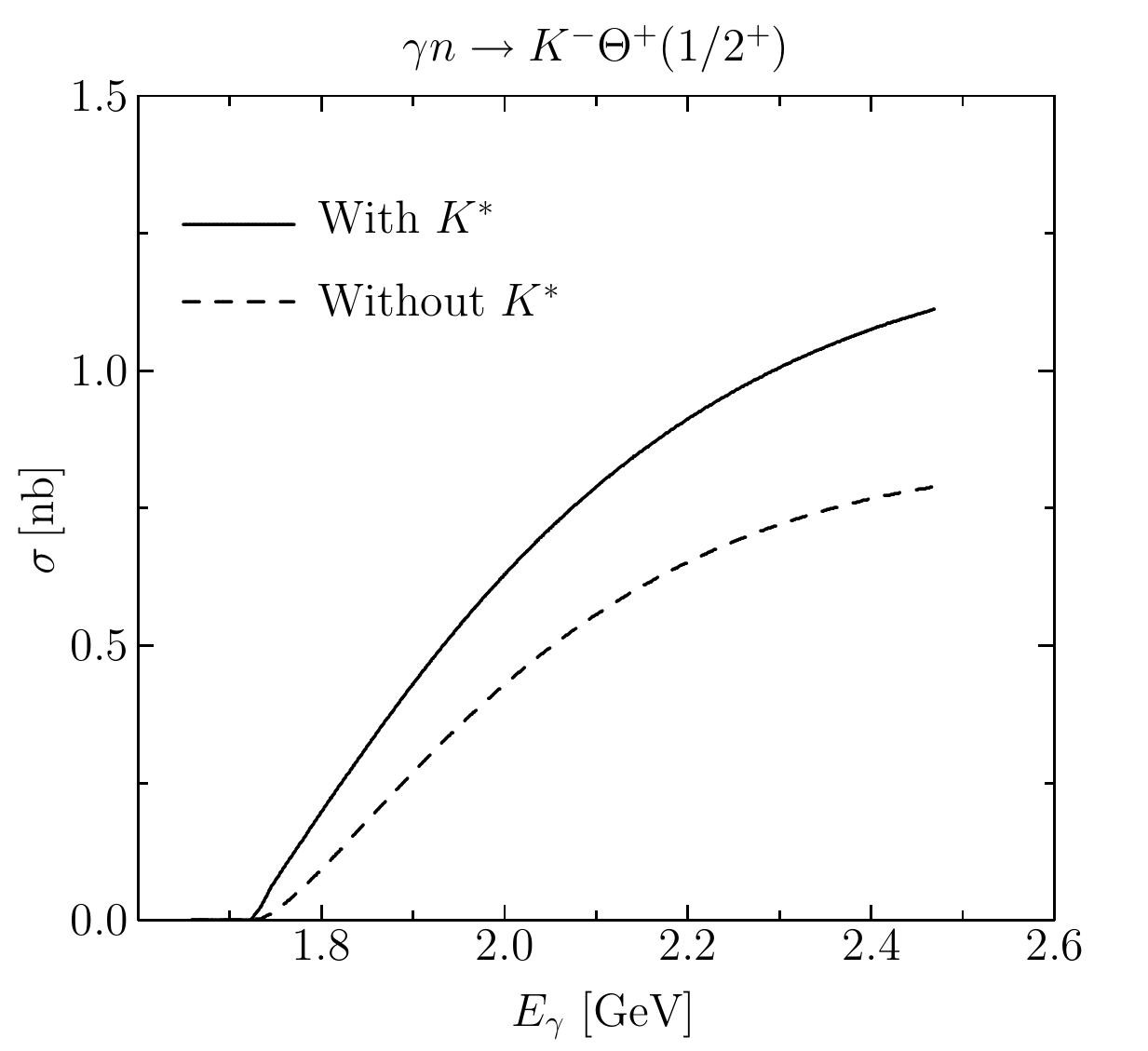} 
\includegraphics[width=7.5cm]{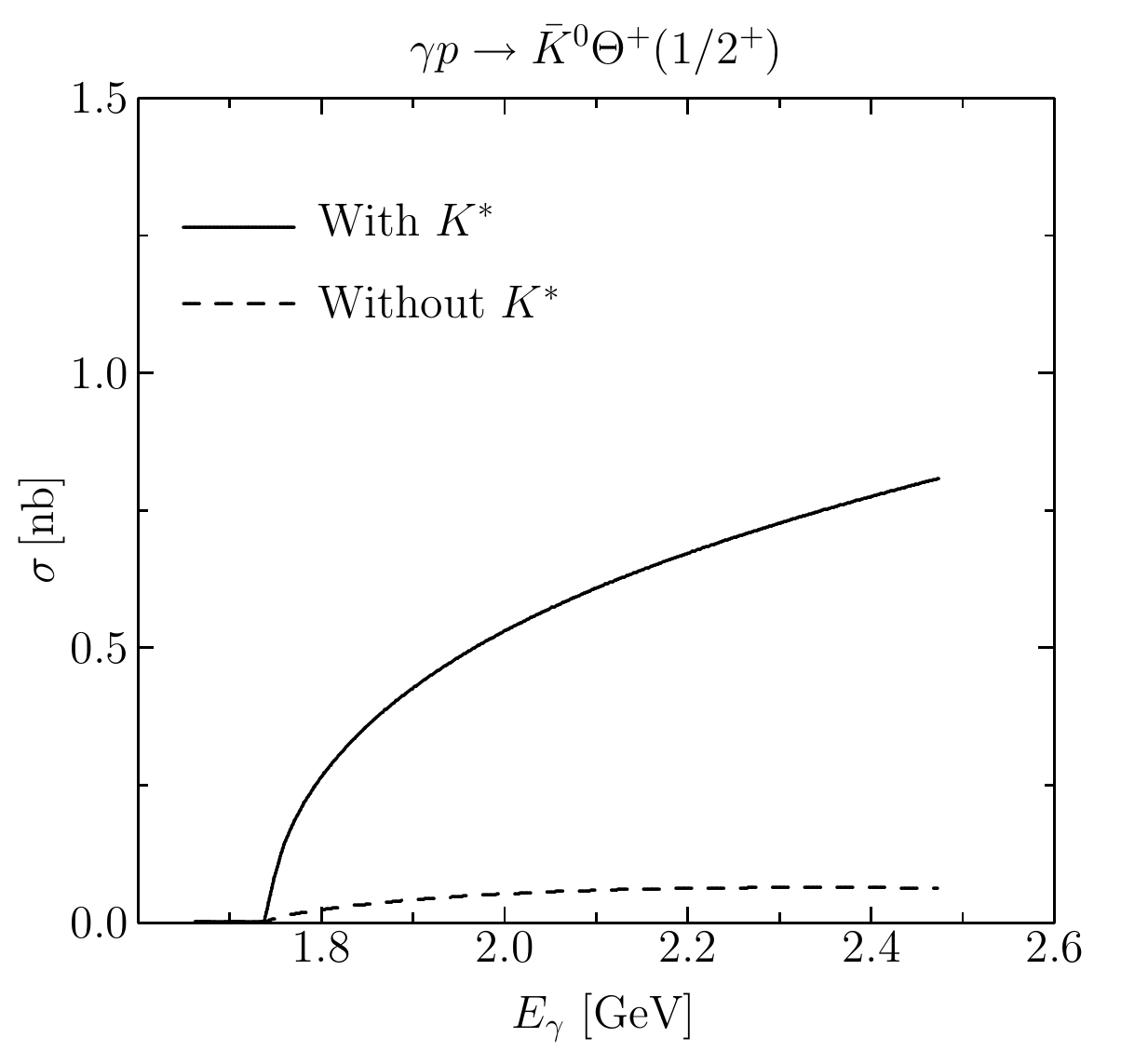}
\caption{Effects of the $K^*$-exchange on the total 
cross sections. The left panels represent those for the $\gamma n \to
K^- \Theta^+$  reaction, while the right panels those for the $\gamma
p \to \bar{K}^0 \Theta^+$.  The solid curves indicate those with all
contributions, whereas the dashed one those without the
$K^*$-exchange.}
\label{fig1}
\end{figure}
Employing the coupling constants and cutoff masses obtained in
Refs.~\cite{Ledwig:1900ri,Ledwig:2008rw}, observables for the $\gamma
N\to K\Theta^+$ reaction were reexamined, based on an effective
Lagrangian approach.  The spin-parity quantum number of the $\Theta^+$
has been assumed to be $1/2^+$ as predicted by the $\chi$QSM.  In this
Section, we briefly summarize the results of Ref.~\cite{Nam:2009ds}.

In Fig.~\ref{fig1}, we draw the
total cross sections for the $\gamma n \to K^- \Theta^+$ and
$\gamma p \to \bar{K}^0 \Theta^+$ reactions with and without the
$K^*$-exchange contribution, respectively, in the left and right
panels.  The $K^*$ exchange contributes to the total cross section for
the neutron target by about $30\,\%$ whereas for the proton target it
is almost everything.  This is due to the fact that no $K$-exchange
contributes to the $\gamma p\to \bar{K}^0\Theta^+$ reaction.  
The results for the neutron target is in qualitative agreement with
the LEPS data~\cite{Nakano:2008ee}.  

\begin{figure}[t]
\includegraphics[width=7.5cm]{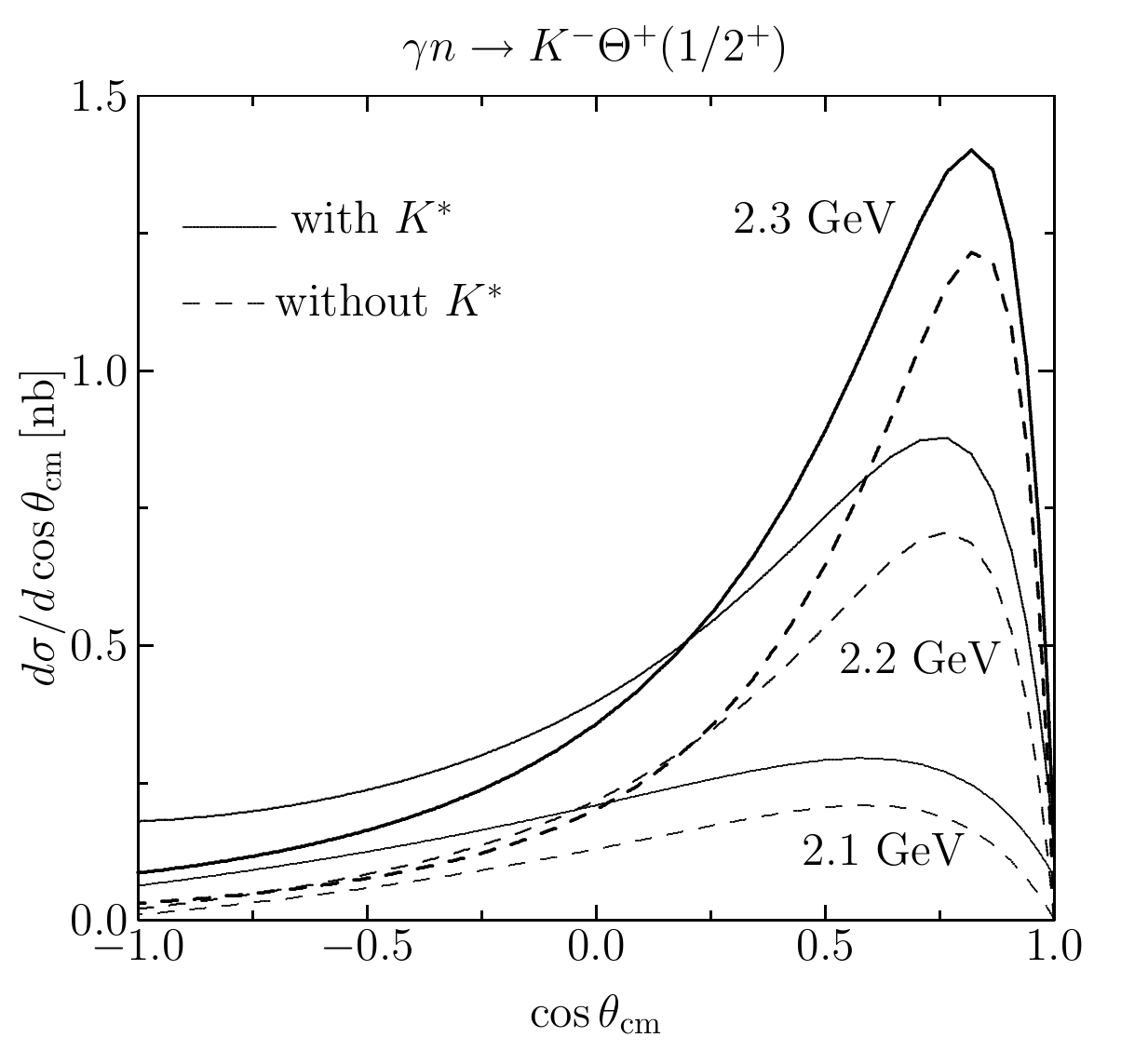}
\includegraphics[width=7.5cm]{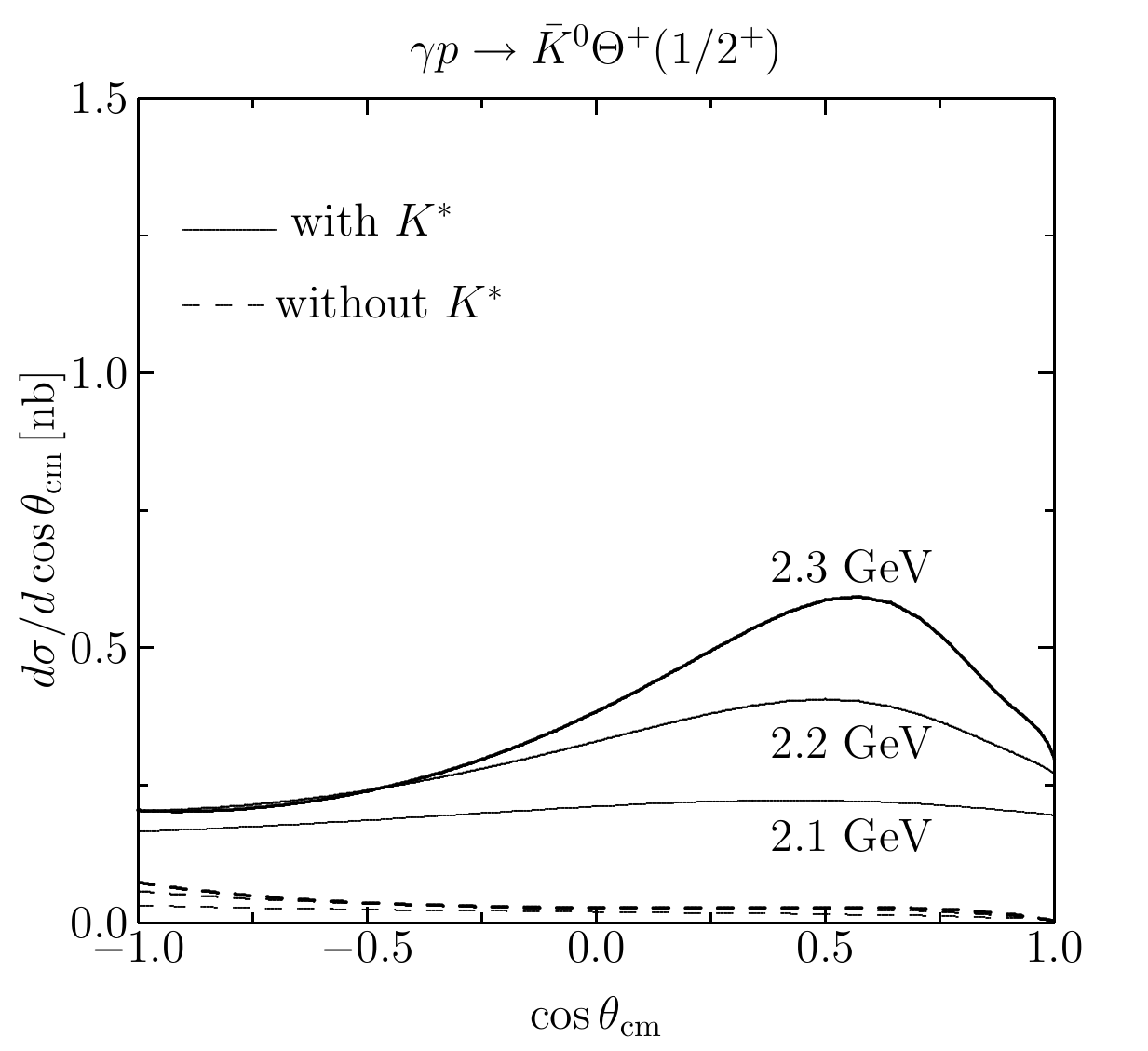}
\caption{Effects of the $K^*$-exchange on the
differential cross sections. The left panels represent those for the
$\gamma n \to K^- \Theta^+$  reaction, while the right panels those
for the $\gamma p \to \bar{K}^0 \Theta^+$.  The solid curves indicate
those with all contributions, whereas the dashed one those without the 
$K^*$-exchange.  The differential cross sections
are drawn for three different photon energies $E_\gamma$, $2.1$ GeV,
$2.2$ GeV, and $2.3$ GeV.}
\label{fig2}
\end{figure}

Figure~\ref{fig2} depicts the differential cross sections for the
$\gamma n\to K^-\Theta^+$ (in the left panel) and $\gamma p\to
\bar{K}^0\Theta^+$ (in the right one) reactions with and without
$K^*$-exchange for three different photon energies 2.1 GeV, 2.2 GeV,
and 2.3 GeV, respectively.  Because of the $K$- and $K^*$-exchange
contributions, the bump structures arise in the region  
$\lesssim60^\circ$ for both the neutron and proton target cases.  As
in the case of the total cross sections, while the $K^*$-exchange
contribution makes the differential cross section about $10\,\%$
enhanced for the neutron target, its effects are remarkably large for 
the proton target. As the photon energy increases, the differential
cross sections also increase consistently, as expected. 

\begin{figure}[t]
\includegraphics[width=7.5cm]{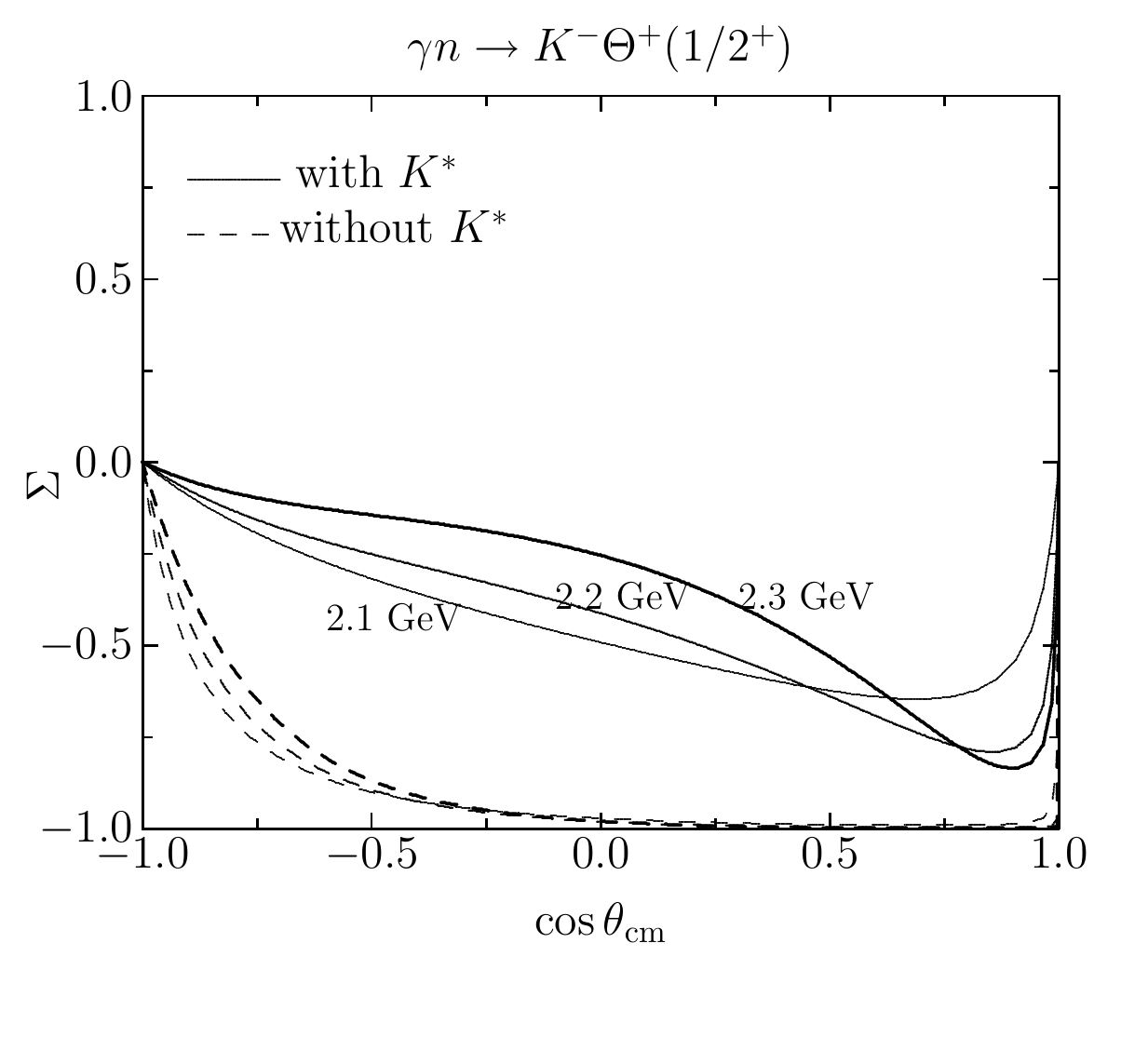}
\includegraphics[width=7.5cm]{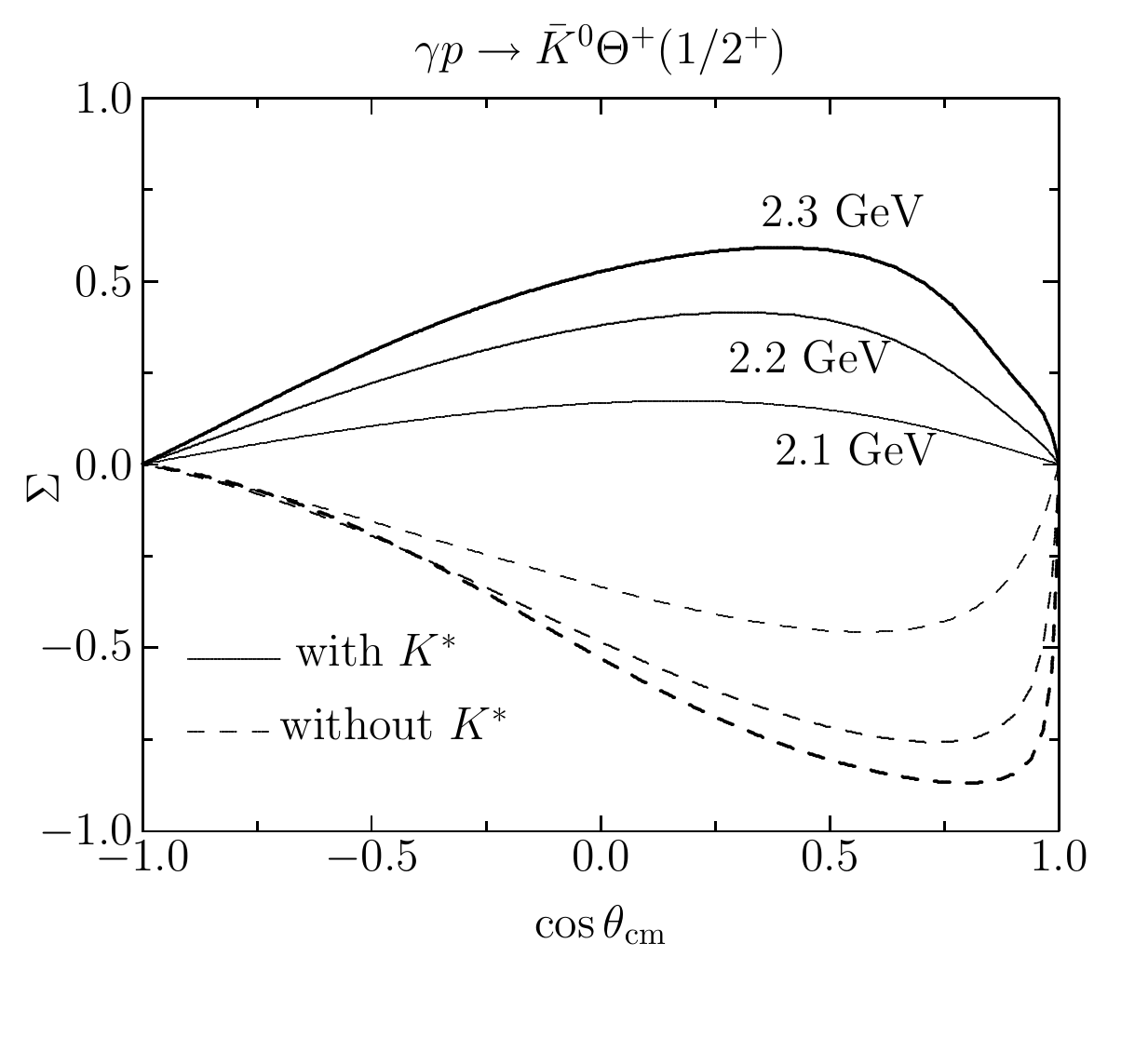}
\caption{Effects of the $K^*$-exchange on the
photon-beam asymmetries. The left panels represent those for the
$\gamma n \to K^- \Theta^+$  reaction, while the right panels those
for the $\gamma p \to \bar{K}^0 \Theta^+$.  The styles are the same as those in Figure~\ref{fig2}}
\label{fig3}
\end{figure}

In the left and right panels of Fig.~\ref{fig3}, we show the
photon beam asymmetries for the $\gamma n \to K^- \Theta^+$ and
$\gamma p \to \bar{K}^0 \Theta^+$ reactions, respectively.  Without 
$K^*$-exchange, the photon beam asymmetry
for the neutron target falls down drastically, starting from the
backward direction, and goes down to almost $\Sigma=-1$ at 
around $\theta_{\mathrm{cm}}=90^\circ$.  It is due to the electric 
meson-baryon coupling of the dominant $K$-exchange contribution.
However, when we switch on the $K^*$-exchange one, the photon beam 
asymmetry decreases mildly from the backward direction to the forward
direction, and then it increases sharply to $\Sigma=0$. On the whole,
the photon beam asymmetry is negative for the neutron target.       
  
In the case of the proton target, $K^*$-exchange shows profound
effects on the photon beam asymmetry. While the photon beam asymmetry 
becomes negative without the $K^*$-exchange contribution, it is
changed into positive values in all the regions with the
$K^*$-exchange contribution considered.  The photon beam 
asymmetry starts to increase from the bakcward direction to the
forward direction, and it gets brought down from around 
$\cos\theta_{\mathrm{cm}}=0.5$.  

%-------------------------------------------------
\section{Summary and conclusions
}
%-------------------------------------------------
In the present talk, we have reviewed recent
works~\cite{Ledwig:1900ri,Ledwig:2008rw} on the $KN\Theta$ 
and $K^* N\Theta$ coupling constants based on the chiral quark-soliton
model. The results are summarized as follows:
the vector and tensor $K^*N\Theta$ coupling constants for the $\Theta^+$:
$g_{K^{*}N\Theta}=0.74 - 0.87$ and $f_{K^{*}N\Theta}=0.53 - 1.16$, and
$\Gamma_{\Theta \to KN}=0.71$ MeV with the $KN\Theta$ coupling
constant $g_{Kn\Theta}=0.83$.  We also discussed the connection of the
presently reported results to the original work by Diakonov et
al.~\cite{Diakonov:1997mm}.  If the present result had used in their
work, they would have obtained $\Gamma_{\Theta \to KN}=3.4$ MeV.
These improvements even solidify the predicted physics in
Ref.~\cite{Diakonov:1997mm} regardless of the criticism by 
Jaffe~\cite{Jaffe:2004qj}. 

Using the coupling constants and cutoff masses obtained in
Refs.~\cite{Ledwig:1900ri,Ledwig:2008rw}, we have reexamined the
photoproduction of the $\Theta^+$.  We found that the results of the
total cross section for the neutron target is compatible with the LEPS
data~\cite{Nakano:2008ee}.

%-------------------------------------------------
\section*{Acknowledgment}
%-------------------------------------------------
This report is prepared for an invited talk given at the Workshop of Excited Nucleon - NSTAR2009 held in Beijing, April $19-22$  2009. H.Ch.K. is grateful to the organizers of the NSTAR2009  for their hospitality. He also wants to express his gratitude to A.~Hosaka, T.~Nakano, and M.~V.~Polyakov for discussions and comments. The work is supported by Inha University Research Grant (INHA-37453). The work of S.i.N. is partially supported by the Grant of NSC\,96-2112-M033-003-MY3 from the National Science Council (NSC) of Taiwan. 
%--------------------------------------------------

\end{document}